\documentclass[a4paper,10pt]{article}
\usepackage[latin1]{inputenc}
\usepackage{cite}
\usepackage{amsmath,amsfonts,amssymb}    
\usepackage{url,xspace}
\usepackage{subfigure}
\usepackage{graphicx}
\usepackage{fullpage}
\begin{document}

\title{Higher Order Statistsics of Stokes Parameters in a Random Birefringent Medium}
\author{by Salem SAID \& Nicolas Le Bihan}
\date{}

\maketitle


\begin{abstract}
We present a new model for the propagation of polarized light in a random birefringent medium. This model is based on a decomposition of the higher order statistics of the reduced Stokes parameters along the irreducible representations of the rotation group. We show how this model allows a detailed description of the propagation, giving analytical expressions for the probability densities of the Mueller matrix and the Stokes vector throughout the propagation. It also allows an exact description of the evolution of averaged quantities, such as the degree of polarization. We will also discuss how this model allows a generalization of the concepts of reduced Stokes parameters and degree of polarization to higher order statistics. We give some notes on how it can be extended to more general random media.
\end{abstract}

\section{Introduction}
The interaction of polarized light with a random medium is of interest to many domains. Examples include imagery, telecommunications, medicine and instrumentation. In this paper, we present a new model for the propagation of a polarized lightwave in a random birefringent medium. This problem is relevant to telecommunications. Indeed, it occurs in optical fibers subject to Polarization Mode Dispersion, or PMD. See for instance~\cite{PMD1,PMD2,PMD3}. The review~\cite{PMDfundamentals} gives a comprehensive introduction to PMD. The approach offered in~\cite{PMD1,PMD2} is based on stochastic differential equations satisfied by the Jones (or Mueller) matrix of a certain length of the medium. 

Our main aim in this paper, is to introduce a new model for the propagation of polarized light in a birefringent random medium. All the main features of this model are generalizable to random media in general. However, we use this simpler situation as a starting point. Unlike the ones given in~\cite{PMD1,PMD2}, the model we present is global and not local. This means that instead of using a stochastic differential equation that describes propagation over short "infinitesimal" distances, we concentrate on the linear operators describing propagation over finite distances. The two approahces are complementary. However, the results that we will describe are more far-reaching than the ones obtained using stochastic diffenretial equations.

One of the main features of this model is that it treats the random medium as a whole. In other words, the random medium is considered as a system and a certain type of statistical relation between its input and output is studied. The microscopic characteristics of the medium are thus related to the parameters of the probability laws involved in the input-output relation characterizing the medium (system). They can be estimated given samples of the input and output of the medium. In fact, the whole probability density of the Mueller matrix of the medium can, in principle, be estimated given samples of the input and output.

The model we will present is based on a decomposition of the higher order statistics of the reduced Stokes parameters along the irreducible representations of the rotation group. The rotation group is important because a birefringent medium acts on the Poincaré sphere by rotations~\cite{BROS}. The relation between the probability densities, on the Poincaré sphere, of the input and output reduced Stokes vectors will be shown to be given by a spherical convolution. The decomposition along irreducible representations of the rotation group is used to obtain from this relation the laws of transformation of the higher order statistics of the reduced Stokes vectors. On the one hand, these laws of transformation constitute generalizations of the Stokes-Mueller formalism to higher order statistics. On the other hand, they will be shown to be a powerful tool for describing the propagation of polarized light in a random medium. This whole approach is actually an example of harmonic analysis on the rotation group~\cite{chirik} which is an instance of non commutative harmonic analysis~\cite{chirik,taylor,Vilenkin}.

Classical models for the state of polarization of a lightwave, such  as the Jones and Stokes models, only involve statistics of order $1$ and $2$~\cite{BROS}. Whenever the fluctuations of the field of the lightwave are non Gaussian, it is necessary to consider higher order statistics. Non Gaussian fields occur in many situations in optics~\cite{dogariu1}. Different ways of including higher order statistics have been proposed. For instance, Réfrégier in~\cite{refregier-kullback,refregier-kullback-1} gives a measure of the degree of polarization based on Kullback relative entropy between the field of the lightwave and a completely isotropic field with the same intensity distribution. In~\cite{luis}, Luis proposes a measure of the degree of polarization based on the mean quadratic distance between the probability density function on the Poincaré sphere of a lightwave and the probability density function of a uniform distribution on the Poincaré sphere. In~\cite{dogariu1} Ellis and Dogariu propose using the correlations of the Stokes parameters to discriminate states of polarization with the same second order statistics. In~\cite{dogariu2}, they use the symmetries of the probability distribution of the Stokes parameters on the Poincaré sphere to make this same distinction. All these models emphasize two aspects. That it is important to include higher order statistics and that this should be done in accordance with the Symmetries of the probability distribution on the Poincaré sphere. Our approach is quite similar to them. It has the additional advantage of formalizing the importance of the spherical symmetry underlying the problem, by using the irreducible representation of the rotation group.

The plan of the paper is the following: In section \ref{problem} we outline the physical situation we wish to consider. In section \ref{representations}, we explain the mathematical tools we wish to use, namely the irreducible representations of the rotation group. In section \ref{main} we give the main equations of our model. In section \ref{stokes}, we discuss how it generalizes the Stokes-Mueller formalism to higher order statistics. In section \ref{propagation} we show how it can be used to describe the propagation of a lightwave in a random birefringent medium. In section \ref{conclusions}, we discuss how our model can be extended to the case of a general random medium and conclude.

\section{Physical problem} \label{problem}
Let us quickly outline the kind of physical setup we are interested in. We consider the effect of a random birefringent medium on the state of polarization of an incident lightwave. We will be interested in two kinds of relations. The first kind of relations is between the input and output states of polarization when the medium is taken as a whole. The second kind is the relations governing the changes in the state of polarization of the lightwave during its propagation in the random medium. The second kind of relations can be considered as a more detailed model for the first. 

In general, the state of polarization of a lightwave will be modelled as a random variable on the Poincaré sphere~\cite{BROS}. If we are considering the input and output states of polarization to a birefringent random medium taken as a whole, we will note $S_{in}$
and $S_{out}$ the random variables on the Poincaré sphere corresponding to the input and output states of polarization. The effect of the random medium on $S_{in}$ is given by its Mueller matrix. Here, it is understood that this Mueller matrix is a matrix-valued random variable~\cite{wolf}. In general, the Mueller matrix acts on the Stokes vector of the incident lightwave and not on the reduced Stokes vector $S_{in}$. However, in the special case of a birefringent medium the Mueller matrix reduces to a rotation matrix acting on the reduced Stokes vector $S_{in}$. According to this discussion, the effect of a random birefringent meduim is given by a random variable $R$ with values in the rotation group $SO(3)$. This effect is given by the following equation:
\begin{equation} \label{randommediawhole}
 S_{out} = RS_{in}
\end{equation}
Where $S_{in}$ and $S_{out}$ are random variables on the Poincaré sphere $S^2$ and $R$ is a $SO(3)$-valued random variable.

When considering the propagation of a lightwave in a birefringent random medium we will give its state of polarization by a stochastic process $S(z)$ with values on $S^2$. Here $z \geq 0$ is the distance along which the wave has propagated through the random medium. For every $z$, $S(z)$ is a random variable on $S^2$ giving the state of polarization of the wave after a distance $z$ of its trajectory in the medium. In relation to equation (\ref{randommediawhole}), we can note $S(0) = S_{in}$ and $S(Z) = S_{out}$ for some given distance $Z$. We make the hypothesis that a length $z$ of any trajectory of the wave can be described using a rotation matrix (as in (\ref{randommediawhole})) noted $R(z)$. We have, as a first model for the propagation:
\begin{equation} \label{randommediapropagation}
 S(z) = R(z)S(0)
\end{equation}
By associating a matrix $R(z)$ to every distance $z$, over which the wave propagates in the medium, we are implicitly ignoring any scattering or beam divergence effects in the medium. This amounts to supposing that light remains collimated in the medium~\cite{poincaretraject}.

In the following section we give the mathematical tools we will use to analyze equations (\ref{randommediawhole}) and (\ref{randommediapropagation}) to obtain the main equations of our model as given in section \ref{main}.

\section{Irreducible representations of the rotation group} \label{representations}
The main idea of this paper is to decompose the higher order statistics of the reduced Stokes parameters along the irreducible representations of $SO(3)$. This decomposition is central to the spherical convolution theorem which we present shortly and which allows us to obtain the laws of transformation of the higher order statistics of the reduced Stokes parameters. These laws of transformation constitute the main equations of our model and are given in section \ref{main}. In fact, we will be interested in the realization of the irreducible representations of $SO(3)$ in the space of square integrable functions on $S^2$. Under this form, the study of the irreducible representations of $SO(3)$ is known as spherical harmonic analysis~\cite{chirik,Vilenkin,taylor}\footnote{See these same references for all the results in this section.}.

Let $f(s) = f(\phi,\theta)$ - where $\phi$ and $\theta$ are the azimuth and polar angle of $s=(s_1,s_2,s_3)$ - be be a square integrable function on $S^2$. $f$ can be decomposed along the orthogonal basis of spherical harmonics $Y^l_m(s)$, where $l \in \mathbb{N}$ and $-l \leq m \leq l$:
\begin{equation} \label{sphericalFT}
 f(s) = \sum_{l \in \mathbb{N}} \sum^{m=l}_{m=-l}(2l+1)\hat{f}^l_mY^l_m(s)
\end{equation}
Where $\hat{f}^l_m$ is the projection:
\begin{equation} \label{invsphericalFT}
 \hat{f}^l_m = \int_{S^2} f(s)\overline{Y^l_m(s)}ds 
\end{equation}
$ds$ is the Haar measure $ds = \sin \phi d\phi d\theta$ and the bar represents complex conjugation. Remeber that spherical harmonics are given by the formula:
\begin{equation} \label{spharmonics-angular}
 Y^l_m(s) = Y^l_m(\phi,\theta) = \sqrt{\frac{(l-m)!}{(l+m)!}} P^l_m(\cos \theta)e^{im\phi}
\end{equation}
Where the $P^l_m$ are associated Legendre functions. And also by the cartesian formula:
\begin{equation} \label{spharmonics-cartesian}
\begin{array}{l}
Y^l_m(s)  = Y^l_m(s_1,s_2,s_3) \\
          = i^m \frac{\sqrt{(l-m)!(l+m)!}}{2\pi l!} \left( \frac{s_1 + i s_2}{\sqrt{s^2_1 + s^2_2}} \right)^m \int^{2\pi}_{0} (s_3 + i
             \sqrt{1-s^2_3}\cos t)^l\cos (mt) dt
\end{array}
\end{equation}
Under the effect of a rotation $r \in SO(3)$, the $(2l+1)$ coefficients $\hat{f}^l_m$ (for every $l \in \mathbb{N}$) transform according to an irreducible unitary representation of dimension $(2l+1)$ of $SO(3)$. In other words, for the rotated function $f_r(s) = f(rs)$ we have the development:
\begin{equation}
 f_r(s) = \sum_{l \in \mathbb{N}} \sum^{m=l}_{m=-l}(2l+1)\hat{\left( f_r \right)}^l_mY^l_m(s)
\end{equation}
Where the coefficients $\hat{\left( f_r \right)}^l_m$ of the development are given (for every $l \in \mathbb{N}$) by the linear transformation:
\begin{equation} \label{irrep1}
 \hat{\left( f_r \right)}^l_m = \sum^{n=l}_{n=-l}\overline{D^l_{mn}(r)}\hat{f}^l_m
\end{equation}
The functions on $SO(3)$, $D^l_{mn}(r)$, which give the elements of the matrix of the linear transformation (\ref{irrep1}) are the matrix elements of the irreducible representation of dimension $(2l+1)$ of $SO(3)$. They can be given explicitely, as functions of the Euler angles $(\phi,\theta,\psi)$ of the rotation $r$:
\begin{equation}
 D^l_{mn}(r) = D^l_{mn}(\phi,\theta,\psi) = e^{-im\phi}P^l_{mn}(\cos \theta)e^{-in\psi}  
\end{equation}
Where the polynomials $P^l_{mn}$ are given by Jacobi polynomials. 

The $(2l+1) \times (2l+1)$ matrices $D^l(r) = \lbrace D^l_{mn}(r) \rbrace$, $-l \leq m,n \leq l$, are unitary and verify the homomorphism property:
\begin{equation}
 D^l(r_1r_2) = D^l(r_1)D^l(r_2)
\end{equation}
Where $r_1,r_2 \in SO(3)$ and $l$ is any natural number.

The functions $D^l_{mn}(r)$, on $SO(3)$, play a similar role to that played by spherical harmonics on $S^2$. Indeed, let $g(r)$ be a square integrable function on $SO(3)$, then $g$ can be decomposed along the orthogonal basis of matrix elements $D^l_{mn}(r)$:
\begin{equation} \label{so3FT}
 g(r) = \sum_{l \in \mathbb{N}} \sum^{m,n = l}_{m,n = -l} (2l+1) \hat{g}^l_{mn}\overline{D^l_{mn}(r)}
\end{equation}
Where $\hat{g}^l_{mn}$ is the projection:
\begin{equation}
 \hat{g}^l_{mn} = \int_{SO(3)}f(r)D^l_{mn}(r)dr 
\end{equation}
and $dr$ is the Haar measure on $SO(3)$, given in terms of Euler angles by $dr = \frac{1}{16\pi^2}\sin \theta d\phi d\theta d\psi$.

It is interesting, at this point, to rewrite the developments (\ref{sphericalFT}) and (\ref{so3FT}) in a matrix form. This will clarify the rest of the article and avoid cumbersome notations. To do this, adopt the following notations: Note $\hat{f}^l$, for every $l \in \mathbb{N}$, the $(2l+1)$ dimensional column vector whose components are the coefficients $\hat{f}^l_m$, $-l \leq m \leq l$, from equation (\ref{sphericalFT}). In the same way, note $Y^l(s)$ the column vector of spherical harmonics $Y^l_m(s)$. Note $\hat{g}^l$ the $(2l+1) \times (2l+1)$ matrix with elements $\hat{g}^l_{mn}$, in equation (\ref{so3FT}). The notation $D^l(r)$ has allready been explained. Using these notations, equation (\ref{sphericalFT}) can be rewritten as follows:
\begin{equation} \label{sphericalFT1}
 f(s) = \sum_{l \in \mathbb{N}}(2l+1)(\hat{f}^l)^{t}Y^l(s)
\end{equation}
Where $t$ stands for transposition. Equation (\ref{so3FT}) can be rewrittin as:
\begin{equation} \label{so3FT1}
 g(r) = \sum_{l \in \mathbb{N}} (2l+1) Tr\left(\hat{g}^l \left( D^l(r) \right)^{\dagger}\right)
\end{equation}
Where $Tr$ stands for the trace and $\dagger$ for the Hermitian conjugate.

As explained above, the spherical convolution theorem will be used to obtain the basic equations of our model. This theorem uses the decompositions (\ref{sphericalFT1}) and (\ref{so3FT1}) to transform a spherical convolution into a family of matrix products. It is an analogue of the classical convolution theorem, which is used to transform a convolution of functions of a real variable into a product of their Fourier transforms. 

The convolution of two functions $g_2$ and $g_1$ on $SO(3)$ is a function $g$ on $SO(3)$ defined as:
\begin{equation} \label{convolution1}
 g(r) = (g_2 * g_1)(r) = \int_{SO(3)} g_2(t)g_1(t^{-1}r)dt
\end{equation}
This definition is analogous to that of the convolution of two functions of a real variable. Formally, it can be obtained from it by replacing the usual $r-t$ by $t^{-1}r$. The convolution of a function $g$ on $SO(3)$ with a function $f$ on $S^2$ is a function $h$ on $S^2$. It has a similar definition:
\begin{equation} \label{convolution2}
 h(s) = (g * f)(s) = \int_{SO(3)}g(t)f(t^{-1}s)dt
\end{equation}
The convolution theorem states that equation (\ref{convolution1}) is equivalent to:
\begin{equation} \label{convolution3}
 \hat{g}^l = \hat{g}_2^l\hat{g}_1^l \mbox{ for } l \in \mathbb{N}
\end{equation}
Where the product on the right hand side is a product of $(2l+1) \times (2l+1)$ matrices. It also states that equation (\ref{convolution2}) is equivalent to: 
\begin{equation} \label{convolution4}
 \hat{h}^l = \hat{g}^l\hat{f}^l \mbox{ for } l \in \mathbb{N}
\end{equation}
Where the product on the right hand side is a product of the $(2l+1) \times (2l+1)$ matrix $\hat{g}^l$ with the $(2l+1)$ dimensional vector $\hat{f}^l$.

\section{Main equations of the model} \label{main}
In this section, we give the main equations of our model. In the following section \ref{stokes}, they will be interpreted as giving the laws of transformation of the higher order statistics of the reduced Stokes parameters. In section \ref{propagation}, they will be used to give a detailed probabilistic description of the propagation of polarized light in a random birefringent medium.

Let us return to the physical situation described by equation (\ref{randommediawhole}) of section \ref{problem}. This equation describes a random medium taken as a whole. It gives the relation between the random variables on the Poincaré sphere, $S_{in}$ and $S_{out}$, describing (respectively) the input and output states of polarization to the medium. In this section, the mathematical tools given in the last section \ref{representations} will be used to analyze equation (\ref{randommediawhole}). First, We will establish the relation, between the probability densities of the random variables $S_{out}$, $R$ and $S_{in}$ appearing in this equation, as a spherical convolution. Then we will use the convolution theorem (\ref{convolution4}) to transform this relation.

Note $p_{in}(s)$ and $p_{out}(s)$ the probability densities of $S_{in}$ and $S_{out}$. Note $p_R(r)$ the probability density of $R$. To see that the relation between these three probability densities is given by a spherical convolution (\ref{convolution2}), apply the law of total probabilities: The probability for $S_{out}$ to take a value near $s \in S^2$, conditionally to the event "$R$ is near $r \in SO(3)$" is equal to $p_{in}(r^{-1}s)$. Indeed, the random variables $S_{in}$ and $R$ are independent (the properties of the medium do not depend on the incident wave). By summing these condition probabilities, we obtain:
\begin{equation} \label{model0}
 p_{out}(s) = \int_{SO(3)}p_R(r)p_{in}(r^{-1}s)dt = (p_R * p_{in})(s)
\end{equation}
Which is a spherical convolution. This relation is analogue to the classical relation stating that the probability density of the sum of two independent real random variables is the convolution of their probability densities~\cite{deconvolution}.

We now apply the convolution theorem - equation (\ref{convolution4}) - to this last relation. Note $\xi^l_{in}$ and $\xi^l_{out}$ the coefficient vectors in the development (\ref{sphericalFT1}) of (respectively) $p_{in}$ and $p_{out}$. Note $R^l$ the coefficient matrices in the development (\ref{so3FT1}) of $p_R$. Then according to (\ref{convolution4}):
\begin{equation} \label{model1}
 \xi^l_{out} = R^l \xi^l_{in}
\end{equation}
This is the first main equation of our model. First of all, it describes the state of polarization of a lightwave using a family of coefficients, \textit{e.g.} $\xi_{in}^l$ $\xi_{out}^l$, instead of a probability density function on $S^2$. It also describes the law of transformation of these coefficients, when the lightwave passes through a random birefringent medium. This description presents some similarities to the model suggested in~\cite{luis}, for the desciption of quantum polarization effects.

Now consider that $S_{in}$ is affected by the composition of two "random elements". That is, let $S_{out} = R_2R_1S_{in}$. Where $R_2$ and $R_1$ are, as in (\ref{randommediawhole}), random variables in $SO(3)$. Let $R = R_2R_1$ and note $p_R(r)$ the probability density of $R$ and similarly note $p_{R_1}$ and $p_{R_2}$. Applying the same reasoning as the one used to obtain equation (\ref{model0}), we have:
\begin{equation} \label{model2}
 p_{R}(r) = \int_{SO(3)}p_{R_2}(t)p_{R_1}(t^{-1}r)dt = (p_{R_2} * p_{R_1})(r) 
\end{equation}
In order to apply the convolution theorem (\ref{convolution3}) to this equation, note $R^l$, $R^l_2$ and $R^l_1$ the coefficient matrices in the development (\ref{so3FT1}) of (respectively) $p_R$, $p_{R_2}$ and $p_{R_1}$. Then, according to (\ref{convolution3}):
\begin{equation} \label{model3}
 R^l = R^l_2R^l_1
\end{equation}
This relation can be generalized to the composition of $n$ random elements, $R = R_nR_{n-1}\ldots R_2R_1$. Using a similar notation to the one in \ref{model3}, we have:
\begin{equation} \label{model4}
 R^l = R^l_nR^l_{n-1}\ldots R^l_2R^l_1
\end{equation}
This is the second main equation of our model. It shows how equation (\ref{model1}) can be applied when the input wave is subjected to the consecutive effect of several random elements. This relation will be used in section \ref{propagation} where we study the propagation of polarized light in random birefrigent media.

\section{Statistical interpretation and generalization of Stokes formalism} \label{stokes} 
In this section, we give a statistical interpretation of the results of the last section, especially equation (\ref{model1}). We study the relation of these results to the Classical Stokes-Mueller formalism~\cite{BROS} as well as to recent works that study the role of higher order statistics in polarization optics~\cite{luis,dogariu1,dogariu2,refregier-kullback}. The main idea is to notice that the coefficient vectors $\xi^l_{in}$ and $\xi^l_{out}$, of equation (\ref{model1}), contain combinations of the moments of order $l$ of the random vectors $S_{in}$ and $S_{out}$, that transform under irreducible representations of $SO(3)$. This observation is used to generalize the notions of reduced Stokes vector and of degree of polarization to higher order statistics.

\subsection{Generalized reduced Stokes vectors} \label{generalizedstokes}
Let us consider a random variable $S$ on the Poincaré sphere, describing the state of polarization of a lightwave. Let $p(s)$ be the probability density of $S$. If $\xi^l$ are the coefficient vectors of the development (\ref{sphericalFT1}) of $p$, then by definition -see equation (\ref{invsphericalFT}):
\begin{equation}
 \xi^l = \int_{S^2}p(s)\overline{Y^l(s)}ds = \mathbb{E} \overline{ \left( Y^l(S) \right) }
\end{equation}
By considering the Cartesian expression (\ref{spharmonics-cartesian}) for the spherical harmonics $Y^l(s)$, it is possible to see that the coefficient vectors $\xi^l$ contain complex combinations of the moments of order $l$ of the vector $S$. Let us take the example of $l = 1$. Using formula (\ref{spharmonics-cartesian}) we can see that the vector $Y^1(S)$ is given by\footnote{$Y^l(s)$ has been defined as a column vector, whence the transpose.} $Y^1(S) = \left( \frac{S_1 - iS_2}{\sqrt{2}}, S_3, -\frac{S_1 + iS_2}{\sqrt{2}} \right)^{t}$. It results from this that $\xi^1 = \left( \frac{\mathbb{E}(S_1 - iS_2)}{\sqrt{2}}, \mathbb{E}(S_3), -\frac{\mathbb{E}(S_1 + iS_2)}{\sqrt{2}} \right)^{t}$. In other words, $\xi^l$ is related by a complex change of basis to the average reduced Stokes vector $\mathbb{E}(S) = (\mathbb{E}(S_1),\mathbb{E}(S_2),\mathbb{E}(S_3))$.

The degree of polarization is classicaly defined using the average reduced Stokes vector. That is, using the second order statistics of the field of the lightwave. It is given by~\cite{BROS}\footnote{For the definition (\ref{DOPold}) to correspond precisely to the usual definition of the degree of polarization, we must add the hypothesis that the total intensity of the lightwave is independent from the other three components of the Stokes vector. This is not a very restrictive hypothesis when considering birefringent media.}:
\begin{equation} \label{DOPold}
 P = \Vert \mathbb{E}(S) \Vert = \sqrt{\mathbb{E}^2(S_1) + \mathbb{E}^2(S_2) + \mathbb{E}^2(S_3)}
\end{equation}
Note that we can also write $P = \Vert \xi^1 \Vert = \sqrt{|\xi^1_{-1}|^2 + |\xi^1_{0}|^2 + |\xi^1_{1}|^2}$. 

This example shows that using the first coefficient vector $\xi^1$ of the development of the probability density of $S$, we retreive the classical average reduced Stokes vector as well as the classical notion of degree of polarization~\cite{BROS}. To generalize this result to higher order statistics, we construct, for every $l \in \mathbb{N}$, a real version of the $(2l+1)$ dimensional coefficient vector $\xi^l$. This can be done as follows. Define for every $l$ the $(2l+1)$ dimensional real vector $S^l$, as follows:
\begin{equation} 
 S^l = \left\{
\begin{array}{lc}
 S^l_m = \frac{-1}{\sqrt{2}} ((-1)^m\xi^l_{-m} + \xi^l_m) = -\sqrt{2}\Re(\xi^l_m) & \mbox{ for } m > 0 \\
 S^l_0 = \xi^l_0 & \mbox{ for } m = 0 \\
 S^l_m = \frac{-1}{i\sqrt{2}} ((-1)^{-m}\xi^l_{m} - \xi^l_{-m}) = \sqrt{2}\Im(\xi^l_{-m}) & \mbox{ for } m< 0
\end{array}
\right.
\end{equation}
We will call the vector $S^l$ the reduced Stokes vector of order $l$. It contains the moments of order $l$ of the vector $S$. For example:
\begin{equation}
 S^2 = (\sqrt{\frac{3}{2}}\mathbb{E}(2S_1S_2),\sqrt{3}\mathbb{E}(S_3S_2),\mathbb{E}(\frac{3}{2}S^2_3-\frac{1}{2}),\sqrt{3}\mathbb{E}(S_3S_1),\mathbb{E}(\sqrt{\frac{3}{2}}(S^2_2-S^2_1)))^t
\end{equation}
 and, $S^1 = \mathbb{E}(S)$.

The definition of the vectors $S^l$ allows the generalization of the notion of degree of polarization to higher order statistics. Indeed, by analogy with formula (\ref{DOPold}), we can define:
\begin{equation} \label{DOPl}
 P^l = \Vert S^l \Vert = \Vert \xi^l \Vert
\end{equation}
We will call $P^l$ the degree of polarization of order $l$. It is possible to prove that $P^l \in [0,1]$. Indeed, $P^l$ is evidently positive. Note also that:
$$
 P^l = \Vert \xi^l \Vert \leq \sqrt{(2l+1)} |\xi^l_n|
$$
Where $-l \leq n \leq l$ is such that $max_{\lbrace -l \leq m \leq l\rbrace}|\xi^l_m| = |\xi^l_n|$. Now~\cite{Vilenkin,chirik}: 
$$
|\xi^l_n|^2 \leq \int_{S^2} p(s) |Y^l_n(s)|^2 ds \leq \int_{S^2} |Y^l_n(s)|^2 ds = \frac{1}{2l+1}
$$
So that $P^l \leq  1$. For $l=1$, the fact that $P^1 = P \in [0,1]$ is well established~\cite{BROS}. It means that the only physical states of polarization are the ones with $\mathbb{E}(S)$ inside the Poincaré sphere.

\subsection{Examples and relation to other works}
In this section, we give a few examples of how the notions of reduced Stokes vector of order $l$ and degree of polarization of order $l$, introduced in the last subsection \ref{generalizedstokes}, can be used to distinguish states of polarization which are indistinguishable in the framework of classical models for polarization~\cite{BROS}. We also explain how our model is related to other recent works on higher order statistics in polarization optics~\cite{luis,dogariu1,dogariu2,refregier-kullback}.

Remember that -see the introduction- classical models for polarization only use the first and second order statistics of the field of the lightwave~\cite{BROS}. This corresponds to using the average Stokes vector or reduced Stokes vector. This approach is sufficient for Gaussian fields but fails for non Gaussian fields~\cite{dogariu1,dogariu2}. Indeed, considering only the average reduced Stokes vector $\mathbb{E}(S) = S^1$ would lead to identifying states of polarization which have the same average $\mathbb{E}(S)$ but might have different higher order moments of this vector. That is, in the formalism introduced in the last ssubsection \ref{generalizedstokes}, different $S^l$ for $l > 1$.

In~\cite{dogariu2}, the three following states of polarization are studied. In the framework of the classical Stokes formalism, they are all considered to be identical states of polarization corresponding to totally depolarized light. However, they all have different higher order statistics: \textbf{i)} A state of polarization with reduced Stokes vector $S$ distributed uniformly on the Poincaré sphere. \textbf{ii)} A state of polarization with reduced Stokes vector $S$ distributed uniformly on the equator of the Poincaré sphere (only linearly polarized light). \textbf{iii)} A state of polarization with reduced Stokes vector $S$ taking the value $(0,0,1)^t$ with probability $1/2$ and the value $(0,0,-1)^t$ with probability $1/2$ (only left or right circularly polarized light).

In all these three cases $\mathbb{E}(S) = S^1 = (0,0,0)^t$ and $P = P^1 = 0$. So that, in the classical Stokes formalism, they all correspond to the same state of polarization. Using the vectors $S^l$, with $l > 1$, we can see how they are different:
\textbf{i)} For this state, the vector $S^l$ is zero and $P^l = 0$ for all $l \geq 1$. \textbf{ii)} For this state $S^1 = (0,0,0)^t$ and $P^1 = 0$, however $S^2 = (0,0,-1/2,0,0)^t$ and $P^2 = 1/2$. \textbf{iii)} For this state $S^1 = (0,0,0)^t$ and $P^1 = 0$, however $S^2 = (0,0,1,0,0)^t$ and $P^2 = 1$.

It appears from these three examples, in addition to the somewhat evident fact that higher order statistics are necessary when studying non Gaussian fields, that in order to call a state of polarization totally depolarized, it is not sufficient to have $P = 0$. Indeed, the state of polarization can verify $P=0$ but still refer to a specific type of polarization ellipse: only linear polarization (state \textbf{ii)}), or only circular polarization (state \textbf{iii)}). More examples are given in~\cite{dogariu2,dogariu1}. In~\cite{luis} a new definition of the degree of polarization is proposed which takes into account higher order statistics. This definition is based on the following quantity:
\begin{equation}
 D = \int_{S^2} [p(s) - 1]^2 ds
\end{equation}
Where $p(s)$ is the probability density of the reduced stokes vector $S$. $D$ is actually a quadratic measure of the difference between $p(s)$ and a uniform distribution, whose probability density is equal to $1$. The degree of polarization is then defined as~\cite{luis}: 
\begin{equation}
 P = \frac{D}{1+D} \in [0,1]
\end{equation}
This definition is closely related to the quantities $P^l$, which we introducecd in the last subsection \ref{generalizedstokes}. In fact~\cite{Vilenkin,chirik}:
\begin{equation}
 D = \sum_{l \geq 1} (P^l)^2
\end{equation}
It is possible to synthesize the results of our model, with the measure of degree of polarization proposed in~\cite{luis}, by defining a totally depolarized state of polarization as one for which $P^l = 0$ for all $l \geq 1$. If we have $P^l = 0$ only for $1 \leq l \leq L$, then we can say that the state of polarization is depolarized to the order $L$. Classical models only consider depolarization to the order $1$. Also, we should consider a state of polarization to be totally polarized only for $P^l = 1$ for all $l \in \mathbb{N}$. This corresponds to a distribution concentrated at one point on the sphere.

Let us make a final observation, without developing it: The measures of degree of polarization, and the criteria for distinguishing states of polarization with the same second order statistics, proposed in this article and in~\cite{luis,dogariu1,dogariu2} do not take into account the intensity distribution of the lightwave. The definition of degree of polarization proposed by Réfrégier~\cite{refregier-kullback,refregier-kullback-1}, is based on the whole probability distribution of the field of the lightwave, including its intensity distribution. A comparative study of these two general approaches may help clarify the importance of including the intensity distribution in a measure of the degree of polarization.

\section{Evolution of the state of polarization during propagation} \label{propagation}
In this section, we use our main equations, (\ref{model1}) and (\ref{model4}) of section \ref{main}, to study the evolution of the state of polarization of a lightwave, propagating in a random birefringent medium. This problem, as mentioned in the introduction, arises in optical fiber telecommunications~\cite{PMD1,PMD2,PMD3}. By using the model we have introduced in this paper, we will be able to achieve a detailed probabilistic description of the problem at hand. In particular, we will be able to give an exact analytical expression for the probability density on the Poincaré sphere, representing the state of polarization, after any distance of propagation. We will also be able to follow exactly the evolution of different averaged quantities, such as the degree of polarization.

The physical problem we are interested in is the one described by equation (\ref{randommediapropagation}) in section \ref{problem}. We have a lightwave, propagating in a random birefringent medium. We note $z \geq 0$ the distance along which the wave has propagated. To each $z \geq 0$ a random variable $S(z)$ on the Poincaré sphere is associated. It represents the state of polarization after a distance $z$ in the medium. As explained in section \ref{problem}, we suppose that there exists, for every $z$ a rotation $R(z)$ such that:
\begin{equation}
 S(z) = R(z)S(0)
\end{equation}
The evolution of the state of polarization during propagation can then be described by a stochastic process $S(z)$ on the sphere, or a stochastic process $R(z)$ on $SO(3)$. Our description of this evolution is based on the mathematical concept of a \textit{Lévy Process} on $SO(3)$~\cite{levy}, which we introduce in the following subsection \ref{levymod}.

\subsection{Lévy process model} \label{levymod}
We model $R(z)$ as a Lévy process on $SO(3)$. This model reflects a set of simple physical properties of the propagation medium. It is based on the following hypotheses:
\begin{itemize}
 \item \textbf{Independent increments:} For $z_1 < z_2$ we have that $R(z_1)$ and $R(z_2)R^{-1}(z_1)$ are independent. Physically, this means that non overlapping parts of the medium are not coupled.
 \item \textbf{Stationary increments:} For $z_1 < z_2$ we have that $R(z_2)R^{-1}(z_1) = R(z_2 - z_1)$. Physically, this means that the medium is homogenous and only locally random. This hypothesis, more generally means that:
\begin{equation} \label{levy}
 R(z) = R(z-z_n)R(z_n - z_{n-1})...R(z_2 - z_1)R(z_1) \mbox{ for } z > z_n > z_{n-1} > ... > z_2 > z_1
\end{equation}
 \item \textbf{Stochastic continuity:} The stochastic process $R(z)$ is stochastically continuous. This means that the probability for $R(z_1)$ and $R(z_2)$ to be different tends to zero as $z_2 - z_1$ goes to zero. Physically, this means that a very short length of the medium can not induce a big change in the state of polarization.\footnote{This does not impose that the trajectories on the Poincaré sphere of $s(z)$ are continuous. They can have jump discontinuities.}
 \item We add the simplifying hypothesis that $R(0) = I$, where $I$ is the $3 \times 3$ identity matrix.
\end{itemize}
A stochastic process on $SO(3)$ -or any other Lie group- verifying these properties is called a left Lévy process, or just a Lévy process~\cite{levy}.

This model can be very effectively reduced, using the main equations (\ref{model1}) and (\ref{model4}). Note $p_{R(z)}(r)$, the probability density of $R(z)$. According to equation (\ref{levy}):
\begin{equation}
 R(z) = R(z-z_1)R(z_1) \mbox{ for } z > z_1
\end{equation}
Using the results of section \ref{main} -equation (\ref{model2})- we can write:
\begin{equation}
 p_{R(z)}(r) = (p_{R(z-z_1)} * p_{R(z_1)})(r)
\end{equation}
Using equation (\ref{model3}), we transform this last equation:
\begin{equation} \label{levyexp}
 R^l(z) = R^l(z - z_1)R^l(z_1)
\end{equation}
Where $R^l(z)$ are the coefficient matrices in the development (\ref{so3FT1}) of $p_{R(z)}$. 

The stochastic continuity of the process $R(z)$ implies the continuity in $z$ of the matrices $R^l(z)$. The only continuous solution of (\ref{levyexp}) verifying $R(0) = I$ is~\cite{levy}:
\begin{equation} \label{generator1}
\begin{array}{l}
R^0(z) = 1 \\ 
R^l(z) = e^{t^lz} \mbox{ for } l \geq 1
\end{array}
\end{equation}
Where $t^l$ (for every $l \geq 1$) is a constant matrix (not function of $z$). These constant matrices are called generators of the process $R(z)$. We have:
\begin{equation} \label{generator2}
 t^l = \frac{d}{dz}R^l(z)\vert_{z=0}
\end{equation}
It follows from the decomposition formula (\ref{so3FT1}) that:
\begin{equation} \label{levyrot}
p_{R(z)}(r) = 1 + \sum_{l\geq 1}(2l+1)Tr(e^{t^lz}(D^l(r))^\dagger)
\end{equation}
Which gives the probability density of $R(z)$ fr any $z$.

From this last result, the probability density on $S^2$ of $S(z)$ can be derived in a direct way. We have already noted that $S(z) = R(z)S(0)$. If $p_{S(z)}(s)$ is the probability density of $S(z)$ and $\xi^l(z)$ are the cefficient vectors in its development (\ref{sphericalFT1}, then by equations (\ref{model0}) and (\ref{model1}):
\begin{equation}  \label{generator3}
\begin{array}{l} 
\xi^0(z) = 1 \\
\xi^l(z) = e^{t^lz}\xi^l(0) \mbox{ for } l \geq 1
\end{array}
\end{equation}
Using the decomposition formula (\ref{sphericalFT1}) it follows that:
\begin{equation} \label{levysphere}
 p_{S(z)}(s) = 1 + \sum_{l \geq 1} (2l+1) (e^{t^lz}\xi^l(0))^{t}Y^l(s)
\end{equation}
Which gives the probability density of $S(z)$ for any $z$.

By modelling the evolution of the state of polarization, during propagation in a random medium, as a Lévy process, the descriptions (\ref{levyrot}) and (\ref{levysphere}) of this evolution have been achieved. Practically, these descriptions give the evolution of the state of polarization in function of the generator matrices $t^l$, $(l \geq 1)$. It is clear from equation (\ref{generator2}) that these matrices characterize the propagation medium locally. That is, they describe propagation over small "ifinitesimal" distances. These matrices are not known \textit{a priori}. There are two ways of giving them: The first way is to use a local model for the propagation, in the form of a stochastic differential equation. Such local models can be found in~\cite{PMD1,PMD2}. The second way is to note that these matrices are parameters of the probability density of $p_{S(z)}$. It is possible to estimate them given realizations of $S(z)$ -see discussion in the introduction. We will return to these two approaches in subsection \ref{estimation}.

\subsection{Depolarization} \label{depolarization}
The most important effect of a random medium on the state of polarization of a lightwave is to depolarize it. After a long distance of propagation, we can expect the state of polarization of the lightwave to become totally depolarized. Here, we use the mathematical description presented in the last subsection \ref{levymod} to give the evolution, during propagation in a random birefringent medium, of the degree of polarization. We see that this evolution tends to a totally depolarized state independently of initial conditions.

It is possible to show, under very general conditions~\cite{levy}, that the (real parts of the) eigenvalues of the generator matrices $t^l$ in equation (\ref{generator1}) are all negative. Since $R^l = e^{t^lz}$, this means that $R^l \rightarrow 0$ exponentially for large $z$. By taking the limit of equation (\ref{levysphere}) for large $z$, we find that $p_{S(z)}(s) = 1$ for large $z$. In other words the probability distribution of $S(z)$ tends to a uniform distribution on the Poincaré sphere, which is characteristic of a totally depolarized state. Note, from equation (\ref{levysphere}), that as the matrices $R^l$ tend to zero $p_{S(z)}(s)$ will tend to a uniform distribution independently of the initial distribution $p_{S(0)}(s)$.

We have established that the state of polarization of a lightwave propagating in a random medium tends, with the distance of propagation, to a totally depolarized state represented by a uniform probability distribution on the Poincaré sphere. Let us now examin the dependence on $z$ of the degree of polarization of order $l$, $P^l$, for $l \geq 1$. Remember that $P^l$ was defined in section \ref{generalizedstokes}, equation (\ref{DOPl}), as the norm of the complex vector $\xi^l$. Using equation (\ref{generator3}), we can write:
\begin{equation}
 P^l(z) = \Vert \xi^l(z) \Vert = \Vert e^{t^lz}\xi^l(0) \Vert
\end{equation}
It is clear in this equation that $P^l(z)$ tends to zero independently of initial conditions. This is, in particular, true for the usual degree of polarization $P = P^1$.

We have just given the law of evolution, \textit{i.e.} the dependence on $z$, of the degree of polarization of order $l$, $P^l$, for all $l \geq 1$. These quantities are an example of what might be called averaged quantities associated to the process $S(z)$. That is, combinations of the averages of functions of $S(z)$. Averaged quantitied are, of course, deterministic. Let $f(s)$ be any real square integrable function on the sphere $S^2$. An example of an averaged quantity is $\mathbb{E}\left( f(S(z)) \right)$. The fact that $f$ is real can be used to slightly transform its development (\ref{sphericalFT1}), in the following way:
\begin{equation}
 f(s) = \overline{f(s)} = \sum_{l \in \mathbb{N}}(2l+1)(\hat{f}^l)^{\dagger} \overline{Y^l(s)}
\end{equation}
Using the fact that, equation (\ref{invsphericalFT}), $\xi^l(z) = \mathbb{E}\left( \overline{Y^l(s)} \right)$, it is possible to write:
\begin{equation}
\mathbb{E}\left( f(S(z)) \right) = \sum_{l \in \mathbb{N}} (2l+1) (\hat{f}^l)^{\dagger} \xi^l(z) 
\end{equation}
Or, using the expression (\ref{generator3}) for $\xi^l(z)$:
\begin{equation} \label{averagedquantities}
 \mathbb{E}\left( f(S(z)) \right) = \sum_{l \in \mathbb{N}} (2l+1) (\hat{f}^l)^{\dagger} e^{t^lz}\xi^l(0)
\end{equation}
The last equation (\ref{averagedquantities} shows that the evolution of any averaged quantity can be followed exactly if the generator matrices $t^l$ are known. Averaged quantities include the entropy of the state of poalrization, the average parameters of the ellipse of polarization, or any other attribute of the state of polarization that we may wish to study. Equation (\ref{averagedquantities}) can for instance be used to establish that the informational entropy of the state of polarization is stricly increasing during propagation in a random birefringent medium. Thus, depolarization can be associated with an increasing informational entropy.

\subsection{Estimation of physical parameters: An example} \label{estimation}
In subsection \ref{levymod}, the evolution of $S(z)$ during propagation was modelled as a Lévy process. This model lead to an analytical formula (\ref{levysphere}), for the probability density of $S(z)$, containing the generator matrices $t^l$ ($l \geq 1$) - see equations (\ref{generator1}) and (\ref{generator2})- as free parameters. As mentioned before, these matrices are related to the local properties of the propagation medium. In order to give them concrete expressions or values, two approaches can be used: The first is to use a local physical model for the evolution of $S(z)$. This model can take the form of a stochastic differential equation~\cite{PMD1,PMD2}. The second way is to estimate them from realizations of $S(z)$, since indeed, these matrices appeare as parameters of the probability density of $S(z)$.

A particularly simple case arises when the generator matrices correspond to the stochastic equation proposed in~\cite{PMD2}. This is a stochastic differential equation describing the evolution of the vector on the Poincaré sphere $S(z)$ during propagation in an optical fibre affected by PMD. The main idea of this model is that $S(z)$ rotates on the Poincaré sphere with an angular velocity which is essentially a white noise vector:
\begin{equation}
 \frac{d}{dz}S(z) = \mu W \times S(z)
\end{equation}
Where $\mu$ is a constant, $W$ is a three dimensional white noise vector, $\times$ is the vector (cross) prodcut and the equation is to be understood as a Stratonovich stochastic differential equation~\cite{PMD2}.

This stochastic differential equation is well known in mathematics~\cite{hsu,levy,chirik,perrin}. It describes \textit{Brownian motion} on the sphere $S^2$. The generator matrices for this process are given by~\cite{hsu,levy,chirik,perrin}:
\begin{equation}
 t^l = \frac{-\mu^2}{2}l(l+1)I_l
\end{equation}
Where $I_l$ is the $(2l+1)\times(2l+1)$ identity matrix. By replacing this result in (\ref{levysphere}), it follows that:
\begin{equation} \label{example}
  p_{S(z)}(s) = \sum_{l \in \mathbb{N}} (2l+1) e^{\frac{-\mu^2}{2}l(l+1)z}(\xi^l(0))^{t}Y^l(s)
\end{equation}
This situation is particularly simple. By starting from the local model suggested in~\cite{PMD2}, we arrive at an expression for the probability density of $S(z)$ which depends on only one free parameter, namely $\mu$, instead of having the (infinite) family of matrices $t^l$ as free parameters. 

note also that, in this case, the degree of polarization of order $l$ takes on a simpler form:
\begin{equation} \label{example1}
  P^l(z) = e^{\frac{-\mu^2}{2}l(l+1)z} \Vert \xi^l(0) \Vert
\end{equation}
Now let us see how, in this simple case which only has one parameter in the probability density of $S(z)$, we can use realizations of $S(z)$ to estimate $\mu$. We need to consider the medium as a whole (a closed system). We note $S(0) = S_{in}$ and consider a length $Z$ of the medium so that we can note $S(Z) = S_{out}$. The medium is represented by the rotation $R = R(Z)$, which is the essential part of its Mueller matrix. In an experimental framework $S_{in}$ should be known to us. Here, it is assumed to be a pure state of polarization such that $S_{in}$ takes the value $(0,0,1)^t$ with probability one (left circular polarization). In this case\footnote{This formula was first given by F. Perrin in 1928.}, the probability density of $S_{out}$ is given by formula (\ref{example}):
\begin{equation} \label{example2}
 p_{S_{out}}(s) = \sum_{l \in \mathbb{N}} (2l+1) e^{\frac{-\mu^2}{2}l(l+1)Z}Y^l_0(s)
\end{equation}
Where we have replaced the values of $\xi^l(0)$ corresonding to $S_{in}$. And the degree of polarization is given by- using formula (\ref{example1}):
\begin{equation} \label{example3}
 P = P^1 = e^{-\mu^2 Z}
\end{equation}
Formula (\ref{example2}) gives the probability density function of $S_{out}$. This probability density function contains $\mu^2$ as a parameter. A standard way of estimating $\mu^2$ is, for instance, maximum of liklihood estimation~\cite{kay}. Using formula (\ref{example2}) in order to find an analytical expression of the maximum of liklihood estimator of $\mu^2$ is not a straightforward task. Still, an exact maximum of liklihood estimator of $\mu^2$ can be found numerically in a standard way. When $\mu^2Z$ is small, the maximum of liklihood estimator of $\mu^2$ can be approximated with the following estimator~\cite{fletcher,pennec}: 
\begin{equation}
 \mu^2 \approx \frac{1}{2ZN} \sum^{i=N}_{i=1} \theta^2_i
\end{equation}
The estimator is evaluated from realizations $S_i$ -with $i = 1,2, \ldots, N$- of the random variable $S_{out}$. Here $\theta_i$ is the polar angle of $S_i$. That is, its angular distance, on the Poincaré sphere, from the inial value $S_{in} = (0,0,1)^t$. The estimator is thus based on the emperical mean of the squared angular distance between the initial state $S_{in}$ and each realization of $S_{out}$. An estimator similar to this one is used in~\cite{poincaretraject} for the standard deviation of speckle noise. Another approach to the estimation of $\mu^2$ is to estimate the degree of polarization and use equation (\ref{example3}) to retreive $\mu^2$. According to our definition of the degree of polarization, formula (\ref{DOPold}), $P$ can be estimated as:
\begin{equation}
 P \approx \left \Vert  \frac{1}{N} \sum^{i=N}_{i=1} S_i \right \Vert 
\end{equation}
The degree of polarization can also be estimated from intensity measurements~\cite{dopestimation}.

\section{Conclusions and outlook} \label{conclusions}
This article was aimed at presenting a new model for the propagation of polarized light in random birefringent media. The physical situations to which this model would be applicable are bounded by the hypotheses introduced in section \ref{problem}. This model was intended to be adapted to a detailed statistical treatment of the physical problems it describes. This is done by including higher order statistics of the reduced Stokes vector and by describing the random medium via a certain type of statistical input/output relation. It was argued that this type of relation, as introduced in section \ref{main}, can accomodate a variety of physical models and make it easier to estimate the physical parameters appearing in these models. The model that was presented is based on a decomposition of the higher order statistics of the reduced Stokes vector along the irreducible representations of the rotation group $SO(3)$, which is the group giving the action of a birefringent medium on the Poincaré sphere. In section \ref{generalizedstokes}, this decomposition was used to generalize the notions of reduced Stokes vector and degree of polarization to higher order statistics. This generalization was discussed in relation to recent works studying the role of higher order statistics in polarization optics. In addition to this more theoretical result, the decomposition along irreducible representation was used, section \ref{main}, to give the laws of transformation, \textit{i.e.} the input/output relations, for the higher order statistics of the reduced Stokes vector of a lightwave propagated through a birefringent random medium. These laws of transformation arise mathematically from the spherical convolution theorem.

In section \ref{propagation}, the evolution of the state of polarization of a lightwave propagating in a random birefringent medium was studied. It was modelled using the concept of \textit{Lévy Processes} on the rotation group, see subsection \ref{levymod}. The framework of Lévy processes on the rotation group was used to give an analytical expression of the probability density on the Poincaré sphere, representing the state of polarization after propagation over any distance in the medium. In subsection \ref{depolarization}, the Lévy process model was used to study the depolarization of a lightwave by propagation in a random birefringent medium. An analytic law for the evolution of the degree of polarization under the effect of propagation in the medium was given. It was shown that depolarization takes place independently of initial conditions. In subsection \ref{estimation}, an example of propagation in optical fibres was discussed. In particular, the probability density on the Poincaré sphere, representing the state of polarization of a lightwave propagating in an optical fibre affected by PMD, was given in an analytical form. The estimation of the physical parameters appearing in this prbability law, from observations of the reduced Stokes vector was discussed.

An important question to adress is how the model presented in this paper, which is specific to birefringent random media, can be generalized to any random medium. Such a general random medium acts on the Poincaré sphere by nonlinear transformations. This makes it difficult to model using the Poincaré sphere formalism. However, if we use the complete Stokes formalism (\textit{i.e.} With all the four components of the Stokes vector), then the medium acts on the Stokes vector essentially by Lorentz transformations~\cite{gopalarao,andersonbarakat}. These are, of course, linear transformations. 

In this article, we have deduced the laws of transformation of the higher order statistics of the reduced Stokes vector by using the irreducible representations of the rotation group. In the case of a general random medium, it is possible to do the same for the higher order statistics of the complete Stokes vector. The finite dimensional irreducible representations of the Lorentz group (which contains the rotation group as a subgroup) should then be used. The finite dimensional representations of the Lorentz group are known as spinor representations~\cite{naimark}. All the main features of the model presented in this paper can be, in this way, generalized to any random medium. However, somme additional technical difficulties would arise, since the Lorentz group, unlike the rotation group, is not compact.

The general idea of the model we have presented is to group as much \textit{a priori} knowledge as possible, about polarized light in random birefringent media, in one consistent statistical framework taking into account the higher order statistics of the wave field. In other words, to construct a general signal model for the state of polarization of a lightwave in a random birefringent medium. This signal model would allow the extraction of significant physical information, in a variety of practical problems, using different signal processing techniques such as detection, estimation, filtering, etc. In this paper, we have presented our signal model and given a toy example of how it can be used. We hope to demonstrate the usefuleness of this model by applying it to concrete problems in future works.

{\footnotesize
\bibliographystyle{plain}
\bibliography{shortnote.bib}
}

\end{document}